\documentclass[jgrga]{agu2001}
\usepackage{graphicx}
\authorrunninghead{PENNA \& QUILLEN}
\titlerunninghead{The Decay of Forbush Events}
\authoraddr{Robert F. Penna \& Alice C. Quillen, Department of Physics
and Astronomy, University of Rochester, Rochester, NY 14627,
USA. (rpenna@pas.rochester.edu, aquillen@pas.rochester.edu)}

\begin{document}

% Uncomment the following code (as well as \usepackage{graphicx}
% above) if you need to include images in draft mode
%\setkeys{Gin}{draft=true}

\title{The Decay of Interplanetary Coronal Mass Ejections and Forbush
Decrease Recovery Times}

\author{Robert F. Penna \& Alice C.~Quillen} \affil{Department of
Physics and Astronomy, University of Rochester, Rochester, NY 14627}

%\email{rpenna@pas.rochester.edu, aquillen@pas.rochester.edu}

\begin{abstract}
We investigate the relation between Forbush cosmic ray decrease recovery time and coronal mass ejection transit time between the Sun and Earth. We identify 17 Forbush decreases from ground based neutron count rates between 1978 and 2003 that occur during the same phase in the solar cycle and can be associated with single coronal mass ejections (CMEs) in the SOHO LASCO CME Catalog or previously published reports, and with specific interplanetary coronal mass ejections (ICMEs) crossing the vicinity of Earth. We find an anti-correlation between Forbush recovery times and CME transit time that contradicts the predictions of simple cosmic ray diffusive barrier models. The anti-correlation suggests that the decay rate of ICMEs is anti-correlated with their travel speed. Forbush recovery times range from seven times the transit time for the fastest disturbance to a fifth the Sun-Earth transit time for the slowest. To account for the large range of measured recovery times we infer that the slowest disturbances must decay rapidly with radius whereas the fastest ones must remain strong. The longest recovery times suggest that the fastest disturbances in our sample decayed less rapidly with radius than the ambient solar wind magnetic field strength.

\end{abstract}
% If using draft mode \end{article} must follow the 
%references section.
\begin{article}

\section{Introduction}
 
Forbush decreases are transient depressions in the Galactic cosmic ray intensity which are characterized by a sudden onset, reaching a minimum within about a day, followed by a more gradual recovery phase typically lasting several days. Though originally thought to be
associated with geomagnetic storms \citep{forbush37}, it is now known from spacecraft measurements that Forbush decreases are also observed distant from planets and so are present in the interplanetary medium \citep{webber86,webber2002}. These decreases are most likely produced by perturbations in the interplanetary magnetic field and particle flow which propagate away from the Sun (e.g., \citealt{morrison56,parker63}).

Variations in the local cosmic ray distribution can be predicted from a time dependent model of the transport of Galactic cosmic rays in the heliosphere \citep{parker65}. The reduced flux of Galactic cosmic rays in the vicinity of an interplanetary disturbance could be due to a variety of physical factors, related to different terms in the cosmic ray transport equation \citep{parker65}. An enhanced solar wind speed leads to increased advection, whereas variations in the magnetic field topology, strength or/and irregularities lead to differences in the
diffusion and drift rates. Some models have focused on enhanced drift (e.g., \citealt{cheng90,rana96}) while others have concentrated on diffusive or scattering models (e.g., \citealt{lockwood86,webber86,wibberenz00,chih86,badruddin02}). For an overview, see \citet{leroux}.  Both drift and scattering mechanisms suggest that the magnitude of a Forbush minimum is proportional to the magnetic field strength and irregularities in the associated interplanetary disturbance.

Galactic cosmic ray decreases are often associated with coronal mass
ejections (CMEs) and their associated interplanetary counterparts,
interplanetary coronal mass ejections (ICMEs) \citep{cane96, cane97,
cane00}.  \citet{cane96} studied 30 years of neutron monitor data and
found 86\% of cosmic ray decreases to be attributable to
CMEs. \citet{cane97} have associated CME ejecta with short-term
particle decreases observed by Helios 1 and 2.

The observed depth of a Forbush event is found to depend on one's trajectory through the ICME \citep{cane96}. Since forward shocks are wider than the driving ejecta, it is possible to pass through a shock but not intercept the CME ejecta.  Forbush decreases are generally of lesser magnitude when only the forward shock is present \citep{cane96}.  \citet{cane96} also found that the depth of a Forbush decrease is dependent on the heliolongitude of the active region which ejected the associated CME.  The depth is largest when the associated CME originates near solar meridian, and the vast majority of Forbush
events are caused by CMEs originating within 50 degrees of 0 degrees heliolongitude \citep{cane96}.  Chromospheric events more than 50 degrees from the solar meridian are rarely associated with ICMEs at the Earth \citep{cane03}.

Following the passage of an interplanetary disturbance which causes a
Forbush decrease, the cosmic ray flux slowly recovers to its initial
level. We refer to the timescale over which the cosmic ray flux
recovers as the Forbush recovery time, $t_{recov}$. The Forbush
recovery time could depend on a number of factors, including the decay
rate and speed of the ICME, the angular size of the ICME, and the
properties of the ambient solar wind
\citep{lockwood86,chih86,leroux}. Because the cosmic ray flux at Earth
is dependent upon the strength of the ICME after it passes the Earth,
it may provide a unique way to probe the structure of ICMEs at radial
locations where we lack spacecraft.

The recent availability of data measured from space-born observatories
allows better constraints on many of these factors. \emph{LASCO} on
board \emph{SOHO} has allowed the measurements of CME ejection speed,
acceleration, size and location at the Sun. Recent measurements of
CMEs have been compiled by Seiji Yashiro and Grzegorz Michalek and
made available in the SOHO LASCO CME
Catalog\footnote{http://cdaw.gsfc.nasa.gov/CME\_list/}; for details
see \citet{zhao03,yashiro04}. The \emph{WIND} and \emph{ACE}
spacecraft allow measurements of gas density and magnetic field vector
as a function of time as the ICME crosses the vicinity of Earth. To
compliment the information available on ICMEs are real time
measurements of the Galactic cosmic ray flux as measured from the
ground. For example, the Moscow Neutron Monitor provides online
\footnote{http://cr0.izmiran.rssi.ru/mosc/main.htm} pressure corrected
neutron hourly and per minute counts since 1958.  The rich diversity
of modern data allows for the accurate identification of CME-ICME
pairs and measurement of their Sun to Earth transit times. This
information can be used to constrain models of Forbush decreases.

In this paper we compile a list of Forbush events associated with CMEs
that have well identified ICME counterparts at the location of
Earth. We combine information on these three components to probe the
relation between ICME transit speed, Forbush size and recovery time,
and ICME decay. In Section 2 we describe our procedure for obtaining a
sample of Forbush events with well-associated CME-ICME pairs, and the
data collected regarding these events, including a description of the
procedure used to measure recovery times. In Section 3 we describe our
observations of the Forbush recovery time's dependence on transit
speed.  Section 4 summarizes and discusses models for the observed
correlation.

\section{Forbush decreases and associated CME and ICME Sample}

Our goal is to find a sample of isolated Forbush events with
well-identified corresponding CMEs and ICMEs. Finding a large sample for this study is difficult because it is necessary to find correspondence between three independent sets of observations. Solar events identified at the Sun (e.g. by \emph{SOHO}) must be related to ICMEs detected at Earth (e.g. by \emph{ACE}, \emph{WIND}, or \emph{GOES}). Our CME-ICME associations were taken from published studies by \citet{cane96}, \citet{sheeley85}, \citet{lindsay99}, and \citet{fry03}.  These events must in turn correspond with Forbush events detected by ground based neutron or/and muon monitors. The Forbush decreases must be sufficiently deep and isolated to allow a good measurement of the recovery time.

A primary reason for the relatively small size of our sample is that we only selected events for which the Sun-Earth link has been verified by several independent parameters. For Events 1--7 of our sample, the CME-ICME pairs were identified and associated with Forbush events by \citet{cane96}. They identified ICMEs primarily based on in-situ measurements of plasma temperature and models relating the plasma temperature to the solar wind flow speed \citep{richardson95}. For these seven ICMEs, they found the associated solar event (i.e., a fast, massive CME) based on the rapid onset of solar particles at the time of a solar flare event. 

For Events 8--12, 14, and 15, the CME-ICME associations were first reported by \citet{gopalswamy01}.  They used \emph{WIND} spacecraft data to identify ICMEs and then examined \emph{SOHO LASCO} images of all CMEs from up to 5 days before the ICME arrival. \emph{SOHO/EIT}, \emph{Yohkok/SXT}, and optical observations were used to eliminate the backside events (\citet{gopalswamy00}).  This allowed identification of a unique source CME for each ICME.

For events 13 and 16, the associations were made by \citet{cho03}. They applied an ensemble of shock propagation models to CMEs, and found a unique ICME in the threshold window for each CME. 

For Event 17, the CME-ICME association was recently reported by \citet{bieber05} with high confidence, based on X-ray and energetic particle data from multiple spacecraft and neutron monitors throughout the time of the event.

\begin{figure*}[h]
\centerline{\noindent\includegraphics[width=31pc]{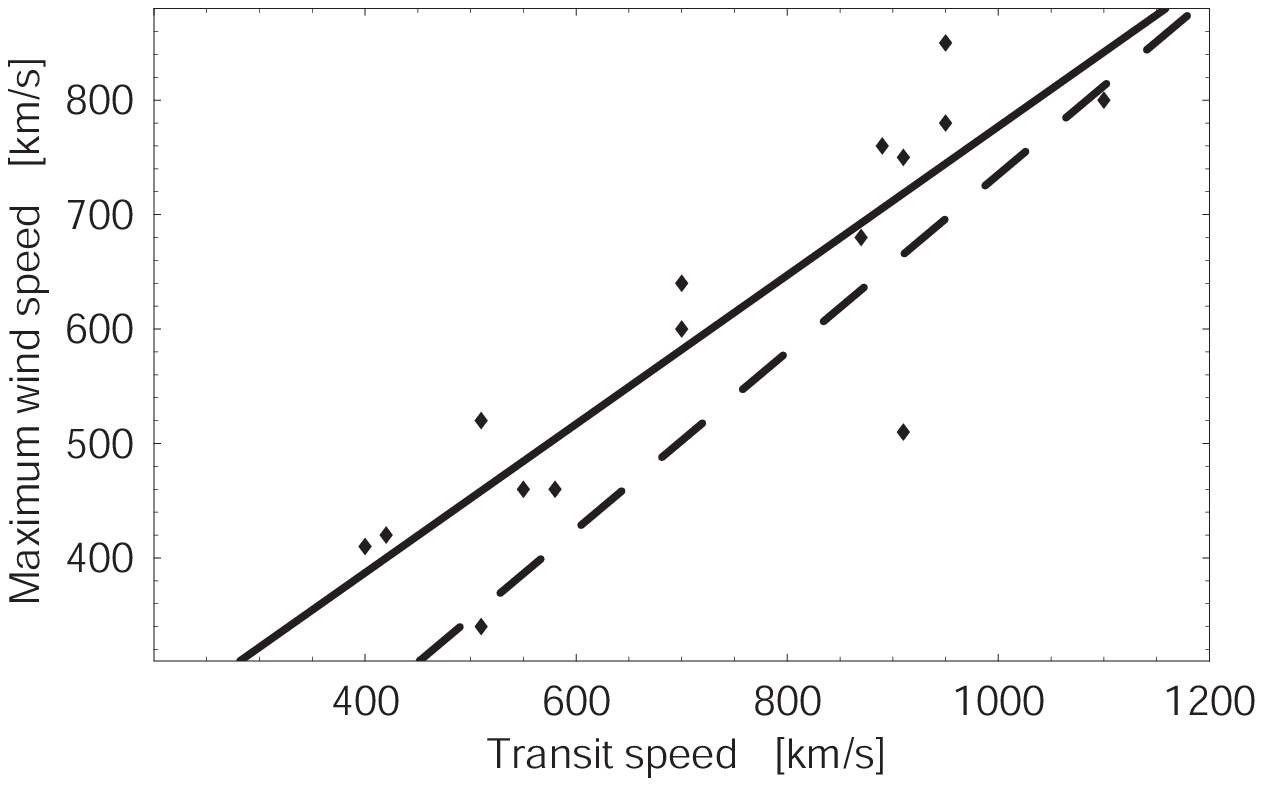}}
\caption{Maximum solar wind flow speed versus shock transit speed for Events 1 and 3--16. Solar wind speed data is not available for Events 2 and 17.  The solid line is the least squares fit to the data.  The dashed line is the empirical relation obtained by \citet{cliver90}.  The close agreement of the data and the empirical relation supports the validity of the CME-ICME associations.} \label{fig:cliveretal}
\end{figure*}

\citet{cliver90} obtained an empirical relation between the Sun-Earth transit speed of interplanetary disturbances and the maximum solar wind flow speed at the time of the ICME at 1 A.U.  In Figure \ref{fig:cliveretal}, we compare the transit speeds and solar wind speeds for Events 1 and 3-16 with the empirical prediction of \citet{cliver90}.  Solar wind speeds were not available for Events 2 and 17.  All of the events plotted in Figure \ref{fig:cliveretal} fall within the scatter of the \citet{cliver90} relation.  The agreement of our sample with the \citet{cliver90} data strengthens our confidence in the CME-ICME associations.  Solar wind speed measurements are from the \emph{IMP8} spacecraft (Events 1, 3--9, and 11--14), or when \emph{IMP8} data is unavailable, from \emph{WIND} (Event 10) and \emph{ACE} (Events 12, 15, and 16).

As a result, we can find only from zero to four Forbush events each year suitable for analysis.  We limited our data to a particular phase in the solar cycle, specifically, we searched for events from three year periods at the start of maxima in the 11-year solar cycle, from 1997--2000, 1986--1989, and 1975--1978.  The solar cycle modulates the GCR distribution, perturbs the heliosphere, and causes structural changes in CMEs.  For example, during solar maximum the average solar wind speed is lower and the average kinetic energy flux in the solar wind is as much as 80\% higher \citep{kallenrode}. \citet{cane03} has shown that on average, the magnetic fields of ICMEs have a more well-organized structure at solar minimum.

To identify a suitable sample of events we studied the set of 57
ICME-CME pairs from December 24, 1996-October 9, 2000, identified in previous studies by \citet{gopalswamy01} and \citet{cho03} and added to this a subset of the list tabulated by \citet{cane96} containing 180 Forbush events occurring during the years 1964--1994.

We investigated the 57 ICME-CME pairs identified by
\citet{gopalswamy01} and \citet{cho03} for associated Forbush
events. We discarded those ICME-CME pairs that were not coincident with Forbush decreases in the neutron count rates measured by the Moscow Neutron Monitor.  We used magnetic field and solar wind measurements from the \emph{ACE} spacecraft to guarantee that the ICME was separated from other significant disturbances by more than 24 hours. When a Forbush event was coincident with arrival of the ICME, we required that the Forbush event also be isolated in time.  This criterion nearly always applied since there are typically only 1--2 significant (depth $>1.5$\%) Forbush events each month.  When overlaps do occur they are clearly visible in the neutron monitor data and we are confident they have not been introduced into our sample.

The arrival times of ICME shocks at Earth are known to high accuracy
(with uncertainty less than 2 hours).  When an ICME results in a
Forbush event, the Forbush event commences within $\sim 10$ hours of
the ICME's arrival at Earth \citep{cane96}.  Since this is much
shorter than the timescale between occurrences of ICMEs and Forbush
events, we have a high level of confidence in the validity of our
associations of individual ICMEs and Forbush events.  If the Forbush
decrease is extremely shallow compared to diurnal variations then
measuring the recovery time becomes difficult. Consequently we
discarded from our study Forbush events with decrease depths less than
1.5\%.

Of the 57 ICME-CME pairs identified by \citet{gopalswamy01} and
\citet{cho03}, only 22 were found to coincide with Forbush
decreases. Only 9 of these Forbush decreases are above 1.5\% in depth
and sufficiently isolated in time.  These 9 events are listed in Table
1 (Events 8--16).

To expand our sample we added seven events from the list of cosmic ray
decreases compiled by \citet{cane96}.  We searched this list for
Forbush events satisfying our suitability criterion over the years
1975--1978 and 1986--1989, years which correspond to the same phase in
the solar cycle as the period 1997--2000 investigated by \citet{cho03}
and \citet{gopalswamy01}. Out of the 12 cosmic ray decreases listed
for these years, we found seven acceptable Forbush decreases.

In addition to this sample we have added the great ``Halloween''
Forbush event of October 29, 2003 because of its extreme properties.

To distinguish between the different parts of each CME-ICME-Forbush
event, the information in Table 1 is separated into 3 groups. For Events
1--16 the first appearance times of the CMEs are those reported by \citet{gopalswamy01}, \citet{cho03}, and \citet{cane96} based on either the \emph{LASCO} coronograph aboard SOHO (\citet{gopalswamy01} and \citet{cho03}) or else Solar-Geophysical Data (\citet{cane96}).  For Event 17, the appearance time is taken from the online SOHO LASCO CME Catalog.  Also under this header we include the heliographic coordinates of the solar source region, whenever an accepted position is available.  The source positions are determined by relating the time of a solar flare to the commencement of a low energy ($<$200 MeV) particle event detected by near-Earth spacecraft (see e.g., \citet{cane96}).  Three events in our sample are not associated with a flare or known filament disappearance and thus their source positions are unknown.  In each case where a source position is available, we find the origin is close to disk center; the average distance from the solar equator is 23 deg. and the average solar longitude is 14 deg. This agrees with the fact that two steps (shock plus driver) were observed for each event in our sample (see below).

Data on the associated ICMEs are grouped in the second part of Table
1. We list the arrival time of each ICME at the L1 Lagrange point and
then the net transit time from the Sun to the L1 point
($t_{transit}$). For Events 1--7, the \emph{ACE} and \emph{WIND}
spacecraft were not available to detect the ICME's arrival at L1. Thus
the estimate of their arrival times is taken to be the sudden
commencement of a geomagnetic storm, as reported by
\citet{cane96}. Since these storms often correspond to the arrival of
the ICME shock rather than the ejecta, this can lead to an error of up
to 12 hours in the transit time \citep{gopalswamy03}.  However given
the very good agreement of both our data sets (see Figure 3) the
actually error in the transit times is probably much less than this.
Also, the fact that both shock and ejecta are present for all the
events in our sample (see below), eliminates much of the potential for
inconsistent measurements and further reduces error in $t_{transit}$.

For events 8--16, the ICME L1 arrival times have been measured
directly in previous studies using solar wind and magnetic field
measurements from \emph{ACE} and \emph{WIND} spacecraft at L1
(\citet{gopalswamy01,cho03}).  Recently, \citet{skoug04} determined
the ICME L1 arrival time for Event 17 based on measurements of
counterstreaming suprathermal electrons, low proton temperatures,
enhanced He$^{++}$/H$^+$ density ratios and smooth rotations of the
magnetic field.

\begin{figure*}[h]
\centerline{\noindent\includegraphics[width=31pc]{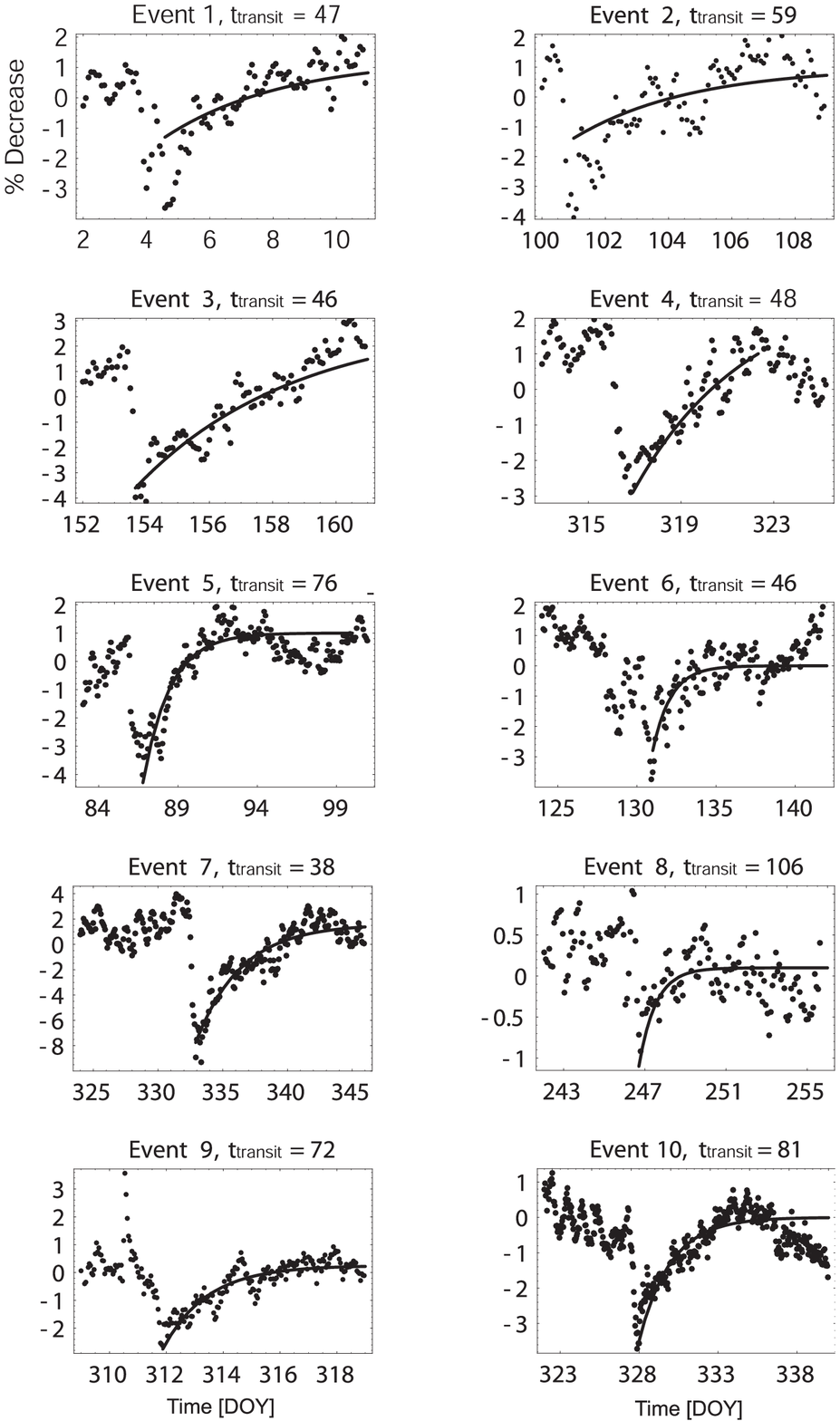}}

\caption{Neutron monitor cosmic ray counts (\% Decrease) as a function of time for Events 1--10 of our sample. The best fit exponentials to the recovery phases of the Forbush events are also shown. As can be seen, Forbush events associated with fast ICMEs (short transit times) tend to recover slowly.} \label{fig:newfig1}
\end{figure*}

\begin{figure*}
\centerline{\noindent\includegraphics[width=31pc]{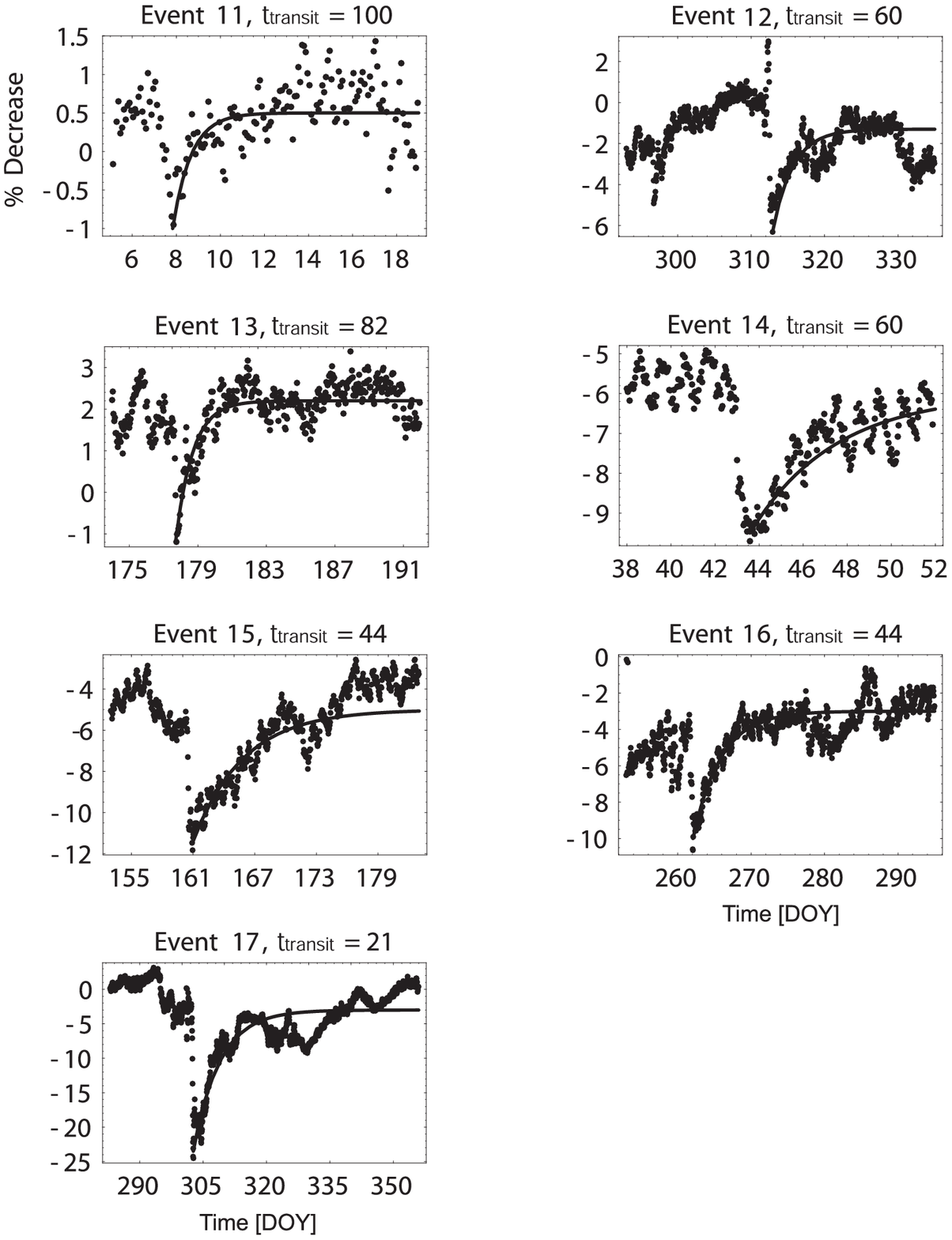}}

\caption{Neutron monitor cosmic ray counts (\% Decrease) as a
function of time for Events 11--17 of our sample. The best fit exponentials to the recovery phases of the Forbush events are also shown. As can be seen, Forbush events associated with fast ICMEs (short transit times) tend to recover slowly.} \label{fig:newfig2}
\end{figure*}

In the last two columns of Table 1 we list the depth and recovery
times of the associated Forbush events. We measured the relative
depths of the Forbush decreases using pressure corrected cosmic ray
count rates from the Moscow Neutron Monitor. This percentage is the
ratio of the minimum cosmic ray flux compared to the average over a
period of a few days preceding the event. The uncertainty in the
measured depth is less than 1\%.  This corresponds approximately to
the level caused by diurnal variations in the count rate. We also
used the neutron count rate to measure the recovery times of the 17
Forbush events listed in Table 1. To do this we fit exponentials of
the form $\exp{(-{t/t_{recov}})}$ to their recovery phases. The best
fitting exponentials to the 17 events in our sample are shown in
Figures \ref{fig:newfig1} and \ref{fig:newfig2}.

Recovery time measurement errors are caused by diurnal variations in
the cosmic ray flux, noise in the count rate, precursors, and slow
increases or decreases in the count rate which lead to differences
between the mean count rate before and after the decrease. To reduce
errors caused by diurnal variations, we removed a characteristic
diurnal variation from the time series before we fit an exponential
to the recovery phase.  We assumed a sinusoidal oscillation with an
amplitude given by the mean diurnal fluctuation during a month of
undisturbed conditions. This month of background data was chosen
separately for each event because the magnitude of the diurnal
variation changes during the year and from year to year. However
%as can be seen from Figure \ref{fig:newfig1},
even after the correction is performed, diurnal effects persist in
many cases. This is in agreement with previous studies which have
found that the amplitude of the diurnal variation is often higher
during periods of enhanced solar activity \citep{lockwood71,
duggal76, nagashima92}. The recovery time of the most unusual data
point, Event 6, is probably incorrect because variations in the
cosmic ray rate before and after this event are about half the size
of the Forbush decrease itself making it difficult to measure the
depth of the event accurately.

To fit the recovery phase we used hourly-averaged cosmic ray counts
for most of the events. To improve the quality of the fit
for Events 1--4, 6--8, and 11 we fit the recovery phase using data
averaged over two hour intervals.

Forbush events are often preceded by strong enhancements in the
cosmic ray anisotropy \citep{lockwood71, duggal76, nagashima92},
which increases the difficultly of measuring the steady pre-event
level. To minimize these uncertainties, we estimated the steady
count rate from the average cosmic intensity over several
undisturbed days preceding and following the Forbush event. Events
10 and 13 occur during prolonged and slow increases in the cosmic
ray flux and so we subtracted baseline count rates from both events
before fitting the recovery phase. In these cases the baseline was
assumed to be a straight line, and was determined from the cosmic
ray count rate using the days preceding and following the Forbush
event.  %As can be seen from Figure \ref{fig:newfig1},
Extremely long recoveries are sometimes interrupted by other
decreases. In such cases only the uninterrupted data has been used
for fitting an exponential.

To estimate the uncertainty of our recovery time measurement, we
artificially created data sets of a decrease with size and noise
variance similar to that of our weakest event, Event 8. We then
applied our fitting procedure to these simulated data sets. Taking
into account our fitting process, we estimate that the net
uncertainty in our measured recovery times $t_{recov}$ is approximately
20\%.   We find that the largest source of error is caused by slow
variations in the mean neutron monitor count rate.

The range of depths (2--21\%) and recovery times (1--7 days) we
measured is consistent with those measured by previous studies,
\citep{lockwood86,cane96}. From Table 1 it is evident that the
Forbush events in our sample with shorter recovery times tend to
have smaller decreases, and shorter transit times. Even for small or
weak decreases of depth $\sim 2\%$ (such as Events 8 and 11 in Table
1), the recovery phases remain nearly exponential and are
distinguishable from noise, diurnal, and slow variations in the mean
count rate. Events with moderate depths, such as Events 1, 5, 10 and 14
which have depths of 4\%, can show either fast ($\sim 27$ hours) or
slow recovery times ($\sim 92$ hours). We infer that the fast
recovery times measured are likely to be real and not an artifact
due to poor measurement of weak Forbush events.

Previous observational studies have shown that the depth of the
Forbush event is dependent on the magnetic field strength
\citep{cane96}. Theoretical studies suggest that the depth is also
dependent on turbulence in the associated ICME, and the ICME's
width. Recovery times on the other hand, are expected to be
independent of event width and magnetic field strength during the
event. Recovery times are expected to only depend on the decay and
speed of the ICME after it passes the Earth (e.g,
\citealt{leroux}). Even though the fast events tend to be stronger
(have larger depths), our sample can be used to probe how Forbush
events recover. We note that the previous studies by \citet{cane96}
found a correlation between transit time and Forbush decrease depth
\citep{cane96}, a correlation which we confirm here. Faster ICMEs
(those with shorter transit times) tend to produce stronger or deeper
Forbush decreases.

It is well known that most Forbush decreases occur in two steps, the
first decrease starting at the shock and the second step occurring
with arrival of the magnetic cloud (e.g. \citet{ifedili},
\citet{cane96}, \cite{lockwood71}).  For all 17 decreases in our
sample both shock and magnetic cloud are present.  For events 1-7
this was known previously (\citet{cane96}, whereas for events 8-17
we have made this determination with solar wind and magnetic field
measurements from the \emph{ACE} Spacecraft.  Recovery time is
measured from the end of the complete two-step decrease.  Since the
two-step effect is the same for all 17 events in our sample it does
not affect recovery time comparisons.

\section{The relation between Forbush decrease recovery time and CME size and ICME transit time}

\begin{figure*}[h]
\noindent \includegraphics[width=40pc]{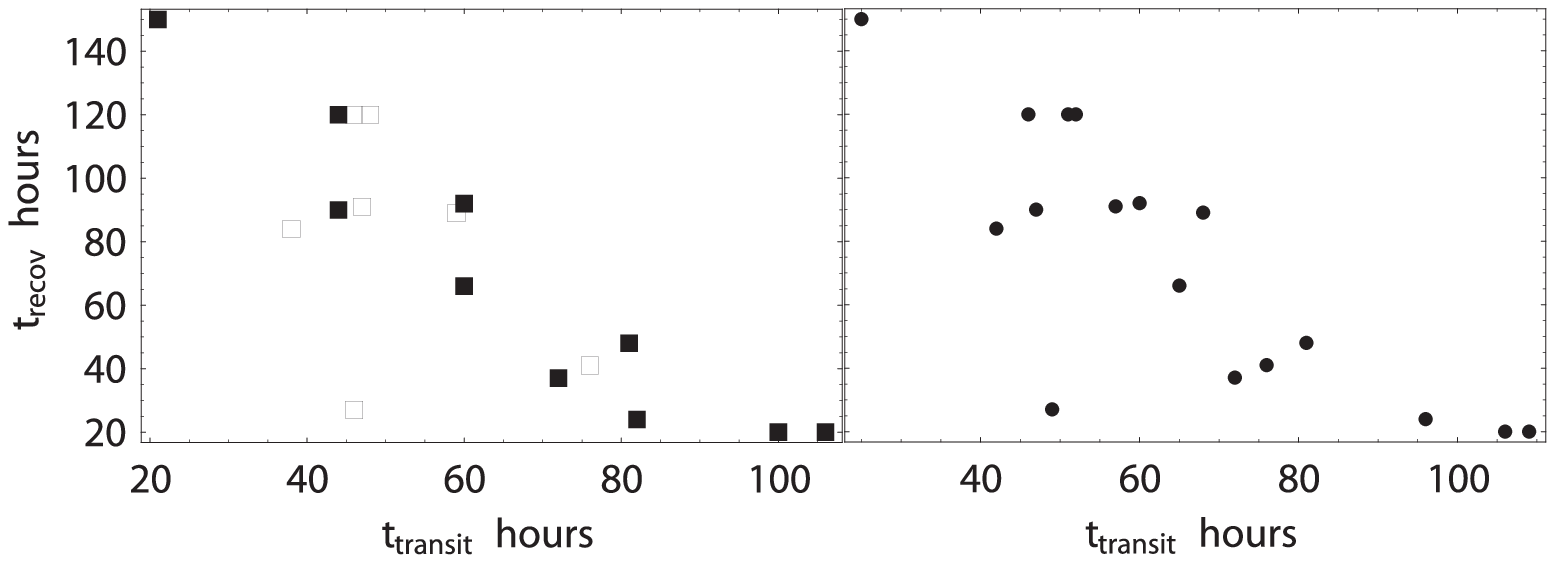}

\caption{Forbush event recovery time versus the transit time of the associated ICME.  In the left panel, solid squares are events for which the ICME arrival time at L1 was known (e.g. \citet{gopalswamy01,cho03,skoug04}) and in this case the transit time has an uncertainty of a few hours (Events 8--17).  White squares are events for which the arrival time of the ICME at L1 is unknown (Events 1--7).  In this case the time of a geomagnetic storm sudden commencement is used instead, and the error is at most 12 hours (\citet{gopalswamy01}).  The close agreement of both data sets suggests the error is in fact much less than this. The storm commencement is a proxy for the shock passage, thus the white squares are biased to the left. In the right panel we took the  arrival time of the ICME to be the onset of the second step Forbush decrease at Earth. 

In both panels, there is a clear anti-correlation between an ICME's velocity and the recovery time of the associated Forbush decrease. The uncertainty in the recovery time is about 20\%. Note that the range of recovery times is large, ranging between 5 times the transit times to 1/4 to 1/5 times the transit times. Models predict that the recovery time should be proportional to the transit time, contrary to the trend shown here.}
\label{fig:rec}
\end{figure*}

Diffusive barrier models for Forbush events suggest that slower
traveling ICMEs will have longer recovery times. The more distant the
diffusive barrier, the more limited effect we would expect on the
Galactic cosmic ray flux (e.g., \citealt{leroux}). In Figure
\ref{fig:rec} we plot the Forbush decrease recovery times versus the
transit times for the events listed in Table 1. Contrary to what we
had expected, fast traveling ICMEs seem to cause Forbush events that
have longer recoveries than those caused by slow traveling ICMEs.  The
Spearman rank coefficient for this correlation is $-0.73$ which gives
a significance level of better than 1\%.

As discussed in Section 2, even though the slower events tend to be
weaker (with shallower Forbush decreases) and so noisier, we are
confident that their recovery times are indeed short.  The recovery
timescales of six events are less than two days long, but the
magnitude of four of these decreases is at least 3\%; large enough
that the measurement of the recovery time is likely to be truly
short, and not a result of measurement errors caused by random
variations in the neutron rate or diurnal variations.

While Figure \ref{fig:rec} shows a statistically significant
anti-correlation between recovery time and transit time, it must be
kept in mind that the fast events in our necessarily small sample
tend to have larger Forbush decreases and higher magnetic fields.
Nevertheless, we do find pairs of events that have similar depths
but very different recovery times.  For example, Events 13 and 14
have similar depths (3.5\%). However Event 13 has a fairly short
recovery time of 24 hours whereas Event 14 has a much longer
recovery time of 92 hours. Even though these two events have similar
depths, their transit times also differ, and the one with the
longest transit time also has the shortest recovery time. Event 13
has a transit time of 82 hours whereas event 14 has a transit time
of 61 hours. Events 12 and 15 have similar depths of 5 and 6.4\%,
and recovery times of 66 and 120 hours, respectively. Again the
largest transit time corresponds to the longest recovery time.

We have displayed Figure \ref{fig:rec} so that Forbush recovery
times can be directly compared to transit times.  We can see from
the figure that the slow events have recovery times that are as
short as 1/5th their transit times, whereas the fastest events have
recovery times that are up to 7 times longer than their transit
times. In the following section we discuss the anti-correlation
shown in Figure \ref{fig:rec} and the range of recovery times
covered in this plot in the context of simple
diffusive barrier models for Forbush events.

\section{Discussion of recovery times}

Within the context of diffusive barrier models, the recovery time,
$t_{recov}$, of a Forbush event could depend on a number of factors
including 1) the decay rate of the propagating disturbance; 2) the
radial gradient of the radial component of the cosmic ray
diffusion coefficient; 3) the velocity of the ICME as it crosses
Earth's orbit, $V_s$; 4) the angular size of the ICME; 5) the
deceleration rate of the ICME after it crosses Earth's orbit. The
recovery time's dependence on the first three factors were
explored and discussed theoretically by \citet{leroux}. One
dimensional analytical advection diffusion models were explored by
\citet{chih86}. Recent work has shown that fast ICMEs decelerate
due to their interaction with the ambient solar wind
\citep{gopalswamy00,gopalswamy01,wang}. However the extent that this
deceleration affects Forbush recovery times has not yet been
explored.

Neglecting drift, but taking into account advection with the solar
wind and diffusion, using a one-dimensional radial approximation,
and neglecting energy changes in the particles, the cosmic ray
transport equation can be approximated by an advection-diffusion
equation
%\begin{equation}
${\partial N \over \partial t} =  {\partial \over \partial r}
   \left( K  {\partial N \over \partial r}
          -V N   \right)
$%\end{equation}
\citep{chih86} where $N(r,t)$ is the number density of the cosmic
ray particles, and $V$ is the speed of the solar wind. We can
consider a propagating magnetic disturbance as a traveling
perturbation in the radial diffusion coefficient $K(r,t)$. The
radial dependence of the diffusion coefficient sets the steady state
solution for the cosmic ray number density. We expect the steady
state value of the number density to be only weakly dependent on
radius at rigidities typical of Galactic cosmic rays responsible for
producing neutrons detected on Earth ($\sim 10$GV). The weak radial
gradient in the number density is consistent with observations of
the cosmic ray flux at somewhat lower rigidities from spacecraft at
different radii in the heliosphere \citep{webb86}.

As discussed by \citet{leroux}, the drop in the cosmic ray flux
during a Forbush event is approximately
\begin{equation}
{\Delta  j \over j} \sim { {V W \over K} {\Delta K \over K} }
\label{fluxdrop}
\end{equation}
where $W$ is the radial width of the disturbance and $\Delta K$ is
the change in the diffusion coefficient during the disturbance.

By assuming that the local cosmic ray flux as a function of time
is related to the drop in the cosmic ray number density at the
location of the disturbance, we can use Equation (\ref{fluxdrop})
to estimate the Forbush recovery time at Earth (e.g.,
\citealt{leroux}). We denote the drop in cosmic ray flux at Earth
(at radius $r_1$) due to an ICME at time $t_1$ by $\Delta
j_1/j_1$.  The recovery time is then approximately $t_{recov} \sim t_2 -
t_1$ where $t_2$ is the time at which the event reaches $r_2$ and
the drop in cosmic ray flux at that radius $\Delta j_2/j_2$ is
$e^{-1} \sim 1/3 $ times that measured at Earth. It is convenient
to define an amplitude $A(t) = {W V} \Delta K/K $ which would
determine the depth of the cosmic ray flux decrease if the ambient
diffusion coefficient were not dependent on radius.  This
amplitude would be constant if the width of the event did not
change in time, and if the diffusion coefficient $\Delta K$ in the
disturbance dropped with radius in the same way as the ambient
diffusion coefficient $K$.

The recovery time can be
estimated from the condition
\begin{equation}
3 A(r_2,t_2)/K(r_2) \approx A(r_1,t_1)/K(r_1). \label{general}
\end{equation}
We expect that the mean diffusion coefficient is inversely
proportional to the mean magnetic field in the solar wind.
Assuming that the magnetic field strength drops with radius as the
solar wind density decreases, we expect $K(r) \propto r^\alpha$.
Voyager observations of the solar wind magnetic field imply that
$B_r \propto r^{-2}$ and variations $\Delta B/|B|$ are nearly
constant. (e.g., \citealt{burlaga98}). From this scaling we expect
$\alpha \sim 2$. We assume a constant velocity for the ICME,
 $r_2  - r_1 \sim  V_s (t_2 - t_1)$,
and an amplitude that drops exponentially with time
$A(t) \propto e^{-t/t_{decay}}$.
We remind the reader that this amplitude depends
on the width of the disturbance, the speed of the solar wind and
the ratio of the change in the diffusion coefficient in the disturbance
divided by the ambient diffusion coefficient.
We have described the decay of the disturbance as a function
of time in terms of one parameter, $t_{decay}$.
Using the above assumptions we find
%\begin{equation}
%t_{recov} = t_2 - t_1 = {r_1 \over V_s} \left({3 A_2 \over A_1} - 1\right).
%\end{equation}
%\begin{equation}
%t_{recov} = t_{transit} \left({3 A_2 \over A_1} - 1\right).
%\end{equation}
\begin{equation}
3 e^{-{t_{recov} \over t_{decay}}} =  \left( 1  + {t_{recov} \over t_{transit}}
                           \right)^{\alpha}
\label{relate}
\end{equation}
where the transit time $t_{transit}  = r_1/V_s$ is the ICME travel
time between the Sun and Earth. This equation relates the observed
quantities shown in Figure \ref{fig:rec},  and
depends only on the decay rate of the ICME ($t_{decay}$)
and on the radial dependence of the diffusion coefficient (set by $\alpha$).
We note that this equation implies that the ratio of the transit time to the
the recovery time is \emph{independent}
of the width of the ICME or its magnetic field strength.
Thus, this ratio should not be related to the depth of the Forbush event.
Consequently even though our sample exhibits a correlation between
ICME transit time and Forbush depth (and so magnetic field strength),
we do not expect a correlation between the ratio of
the recovery time to the transit time.
%Such a correlation would imply that
%the decay rate of the disturbance is related
%to its magnetic field strength.

We now discuss Equation (\ref{relate}) in the context of Figure
\ref{fig:rec}. The slow events shown in Figure \ref{fig:rec} have
recovery times that are well below their transit times.   The
slowest events have transit times of $\sim 100$ hours and recovery
times that are one fourth to one fifth as large as their transit
times. Equation (\ref{relate}) only allows extremely short
recovery times in the limit that the recovery time exceeds the
decay rate of the disturbance; $t_{recov} \gg t_{decay}$.  In this case
the recovery is approximately equivalent to the decay rate; $t_{recov}
\sim t_{decay}$. We conclude that the slow events are likely to be
decaying rapidly.

In contrast, the fast events shown in Figure \ref{fig:rec} have
recovery times that are well above their transit times.  The
fastest event has a recovery time that is approximately seven times
the transit time. However Equation (\ref{relate}) places an upper
limit on the recovery time. The maximum recovery time allowed is
twice the transit time if $\alpha=1$ and equivalent to the transit
time if $\alpha\sim 2$. These limits are reached when the recovery
time is far smaller than the decay rate of the disturbance; $t_{recov}
\ll t_{decay}$. For both cases, the measured recovery times for
the fast events are longer than permissible by Equation
(\ref{relate}).

There are two possible ways to account for the extremely long recovery
times of the fast events.  The events could be decelerating, or the
function $W \Delta K/K$ could be increasing instead of decreasing.  We
first consider the possibility that the ICMEs are
decelerating. Recovery times are longest in the limit that the decay
rate is long. We can use Equation (\ref{general}) to estimate the
recovery time. Assuming that the mean diffusion coefficient $K \propto
r^2$, and that the amplitude of the event ($W \Delta K/K$) is constant
with time, the recovery time is equivalent to the time it takes the
event to propagate between 1 AU and approximately 2 AU. The very
slowest this could occur is the average solar wind speed times 1 AU,
or about 100 hours (assuming a solar wind speed of 300 km/s). This
assumption is extreme because it would require the disturbance to
immediately drop to ambient solar wind velocity just as it passed
Earth even though its rapid transit times implies a much higher
velocity (greater than 1500 km/s). The observed decay times of the
fastest events are uncomfortably near this value or 150 hours long.
For deceleration to account for the long recovery times, we would
require that the event drop to nearly ambient solar wind speeds within
a fraction of an AU. While strong ICMEs do decelerate, even at
distances of the Voyager spacecraft in the outer heliosphere, these
events are still propagating faster than the ambient solar wind
\citep{wang}. The Halloween event was detected by Cassini on Nov 11,
2003 and by Voyager in April 2004.  The transit time between Earth and
Cassini shows that this event was propagating at a speed of $\sim
1000$ km/s at the location of Cassini (9 AU) and 600 km/s at the
location of Voyager (75 AU).  While this event also provides evidence
that the blast wave decelerates, it shows that the deceleration is
fairly gradual.  At a speed of 1500 km/s the disturbance would have
reached 2 AU only 1.5 days after it reached Earth. Taking into account
the empirical interplanetary deceleration relation obtained by
\citet{gopalswamy00}, the disturbance still would have reached 2 AU in
well under two days.  Again, this timescale is significantly shorter
than the observed recovery time, suggesting that the deceleration of
the disturbance is not sufficiently fast to account for the long
recovery times.

While fast ICME's do decelerate, we find that they do not
decelerate fast enough to account for the associated long Forbush
recovery times.  Consequently we must consider alternative
explanations for their long recovery times.  Above we defined an
amplitude function $A(t) = VW\Delta K/K$ which we assumed would be
constant if the disturbance was not decaying.  However this
implicitly assumes that expansion of the disturbance as it travels
causes the magnetic field (and thus the diffusion coefficient) in
the disturbance to decay with the same dependence on radius as is
true for the ambient solar wind.   For a constant $A(t)$ and
reasonable values of $\alpha \sim 1$ or $2$, Equation
(\ref{relate}) does not allow a recovery time above twice the
transit time. This conflicts with the observed recovery times of
the strongest events which can be 5 times larger than the transit
times. To allow such large recovery times, we require $A(t) = W
V\Delta K/K$ to increase with radius with respect to the ambient
solar wind. If we assume that the diffusion coefficient scales
with the mean value of the magnetic field, then an increasing
$A(t)$ implies that the magnetic field in the disturbance times
its width should decrease less rapidly as a function of radius
than the ambient solar wind. Fast CMEs could sweep up larger
shocks from the ambient solar wind which precede the arrival of
their ejecta. The observed deceleration of ICMEs does imply that
energy is lost by the traveling disturbance.

Alternatively fast and strong Forbush events might be associated
with multiple ICMEs which could merge, causing apparently large
recovery times.  This effect may be relevant to Event 17.  The ICME
associated with Event 17 is followed by a second ICME in under a day
(but more than 10 hours) and both cause strong geomagnetic storms,
which are separated by 36 hours. We can associate Event 17 with the
first ICME in the pair because it is coincident with the first
storm.  The second ICME is expected to increase the apparent
recovery time. However because of the extreme size of the observed
recovery we are confident that it is truly long and not merely due
to the combined action of the two ICMEs.  Even after reducing $t_{recov}$
by half it remains above the mean time for our sample.   In reality
the effect of the second ICME is probably much less then a factor of
two because the second ICME is smaller than the first.

\section{Conclusions}

Using ground based neutron counts available from the Moscow neutron
monitor we have searched for Forbush decreases coincident with CMEs
that have been matched to ICMEs by previous studies
\citet{gopalswamy01,cho03,cane96}.  We also added to our sample the
October 29, 2003 event because of its extreme transit time and
recovery time. After discarding CME-ICME pairs lacking Forbush
decreases, decreases of low amplitude and non-isolated decreases, we
obtained a sample of 17 CME and ICMEs matched to observed Forbush
decreases. Our sample exhibits a strong anti-correlation between the
ICME Sun-Earth transit time and the Forbush recovery time as measured
on Earth. This anti-correlation is opposite to the prediction of
simple diffusive barrier models for cosmic ray transport. These models
predict that the Forbush recovery time should be approximately
proportional to the ICME travel speed. However the recovery times that
we measured deviated strongly from this prediction. We found that the
fastest events have recovery times over \emph{seven} times their
transit times, whereas the slowest events have recovery times one
fourth to one fifth times their transit times.

The extreme range of recovery to transit time ratio places strong
constraints on the diffusive barrier models for Forbush decreases.
The short recovery times of the slow events suggest that they
rapidly decay. The long recovery times of the fast events suggest
that their amplitudes might even increase with radius rather than
decrease.  For the fast events, the strength of the ICME could be
increasing as it travels because multiple ICMEs merge or because
the shock preceding the CME ejecta is enhanced as more ambient
solar wind is encountered by the rapidly traveling ICME. These
possible explanations can be investigated with better theoretical
modeling or simulations and by studying the observed physical
properties of ICMEs as they pass through the heliosphere.

Our sample contains a significant bias; the fast (short transit time)
events tend to create larger Forbush decreases and the slow events
tend to cause weaker Forbush decreases.  However diffusive barrier
models imply that the recovery time of a Forbush event is independent
of the depth of the decrease and the strength of its magnetic field,
and is instead primarily dependent on how it decays as it travels
through the heliosphere.  Consequently the bias in the sample does not
account for the anticorrelation between the transit times and recovery
times that we have found here.  A model which incorporates a relation
between the decay rate of the disturbance and travel speed could
explain the anticorrelation found here and would also be consistent
with the observed correlations between Forbush depth and speed
\citep{cane96}.

This work suggests that slow CMEs decay rapidly, in agreement with
near-Sun observations of CMEs by \citet{gopalswamy04}, whereas fast
CMEs remain strong.  As consequence, we expect that slow ICMEs will
not pose serious space weather threats, whereas fast ICMEs will.
Because space weather forecasting is critical to existing and future
space bound missions, we are motivated to further investigate the
relationship between ICME strength and decay and CME ejection speed.

\begin{acknowledgments}
We thank Eric Blackman, Chris St. Cyr, Joe Willie and Kevin
McFarland for helpful discussions. Support for this work was
provided by NSF grants PHY-0242483 (for Rochester's REU program),
AST-0406823 and PHY-0134988, the Research Corporation under Award
number CS0857, the National Aeronautics and Space Administration
under Grant No. NNG04GM12G issued through the Origins of Solar
Systems Program, and a Society of Exploration Geophysicists (SEG)
Schlumberger Scholarship to Robert Penna. The \emph{SOHO LASCO CME
Catalog} is generated and maintained by NASA and The Catholic
University of America in cooperation with the Naval Research
Laboratory. SOHO is a project of international cooperation between
ESA and NASA.  We thank the ACE MAG instrument team and the ACE
Science Center for providing the ACE data. IMP 8 data was received from the MIT Space Plasma Physics Group. Neutron monitors of the
Bartol Research Institute are supported by NSF grant ATM-0000315.

\end{acknowledgments}

%\vfill\eject

%\vfill\eject

\vfill\eject

%%%%%%%%%%%%%%%%%%%%%%%%%%%%%%%%%%%%%%%
%remove table and submit to JGR separately

\begin{table*}[h]
\def\nodata{...}
%\tablecolumns{11} \tabletypesize{\footnotesize} %\tablewidth{0pc}
\caption{CME-ICME pairs and related Forbush Events}
\begin{tabular*}{\textwidth}{@{\extracolsep{\fill}}llllllcllcc}
%{lllcllrcccc}
\hline \cr
 \multicolumn{1}{c}{} &
 \multicolumn{3}{c}{CME} &
 \multicolumn{3}{c}{ICME} &
 \multicolumn{4}{c}{Forbush Event} \cr
%
%\multicolumn{1}{l}{} &
\cline{2-4} \cline{5-7}  \cline{8-11}

\cr \multicolumn{1}{c}{} & \multicolumn{2}{c}{Time} &
\multicolumn{1}{c}{} & \multicolumn{2}{c}{Arrival} &
\multicolumn{1}{c}{} & \multicolumn{2}{c}{Onset} &
\multicolumn{2}{c}{} \cr
\cline {2-3} \cline{5-6} \cline{8-9} \cr \multicolumn{1}{l}{Event} &
\multicolumn{1}{l}{Date} & \multicolumn{1}{l}{UT} &
\multicolumn{1}{l}{Position} & \multicolumn{1}{l}{Date} &
\multicolumn{1}{l}{UT} & \multicolumn{1}{c}{$t_{transit}$} &
\multicolumn{1}{l}{Date} & \multicolumn{1}{l}{UT} &
\multicolumn{1}{c}{Decrease} & \multicolumn{1}{c}{$t_{recov}$} \cr
\multicolumn{3}{c}{} & \multicolumn{1}{c}{deg.} & \multicolumn{2}{c}{}
& \multicolumn{1}{c}{hours} & \multicolumn{2}{c}{} &
\multicolumn{1}{c}{\%} & \multicolumn{1}{c}{hours} \cr \hline
%\colhead{Decrease} & \colhead{$t_0$} \\
%\colhead{} & \colhead{} & \colhead{} & \colhead{(degrees)} & &
%\colhead{} & \colhead{} & \colhead{(hours)} & \colhead{(nT)} &
%\colhead{(\%)} & \colhead{(hours)}}
%

1 & 1978 Jan 1 & 22:00 & 21S 06E & 1978 Jan 3 & 21:00 & 47 & 1978 Jan
  4 & 07:00 & 4 & 91 \cr

2 & 1978 Apr 8 & 02:00 & 19N 11W & 1978 Apr 10& 13:00 & 59 & 1978 Apr
  10 & 22:00 & 6 & 89 \cr

3 & 1978 May 31& 11:00 & 20N 43W & 1978 June 2& 09:00 & 46 & 1978 June
  2 & 15:00 & 6 & 120 \cr

4 & 1978 Nov 10 & 01:00 & 17N 01E& 1978 Nov 12& 01:00 & 48 & 1978 Nov
  12 & 04:00 & 5 & 120 \cr

5 & 1989 Mar 23& 20:00 & 18N 28W & 1989 Mar 27 & 00:00 & 76 & 1989 Mar
  27 & 00:00 & 4 & 41 \cr

6 & 1989 May 5 & 08:00 & 30N 04E & 1989 May 7 & 06:00 & 46 & 1989 May
  7 & 09:00 & 3 & 27 \cr

7 & 1989 Nov 26 & 18:00 & 25N 03W & 1989 Nov 28 & 08:00 & 38 & 1989
 Nov 28 & 12:00 & 9 & 84 \cr

8 & 1997 Aug 30 & 02:00 & 30N 17E & 1997 Sept 3 & 12:00 & 106 & 1997
  Sep 3 & 12:00 & 2 & 20 \cr

9 & 1997 Nov 4 & 06:00 & 14S 33W & 1997 Nov 7 & 06:00 & 72 & 1997 Nov
  7 & 06:00 & 3 & 37 \cr

10 & 1997 Nov 19 & 12:00 & \nodata & 1997 Nov 22 & 21:00 & 81 & 1997
  Nov 22 & 21:00 & 4 & 48 \cr

11 & 1998 Jan 2 & 23:00 & 47N 03W & 1998 Jan 7 & 03:00 & 100 &1998 Jan
   7 & 12:00 & 2 & 20 \cr

12 & 1998 Nov 5 & 21:00 & 22N 18W & 1998 Nov 8 & 09:00 & 60 &1998 Nov
   8 & 14:00 & 5 & 66 \cr

13 & 1999 June 22& 19:00 & \nodata & 1999 June 26 & 05:00 & 82 & 1999
   Jun 26 & 19:00 & 3 & 24 \cr

14 & 2000 Feb 10 & 03:00 & 27N 01E & 2000 Feb 12 & 15:00 & 60 & 2000
   Feb 12 & 15:00 & 4 & 92 \cr

15 & 2000 June 6 & 16:00 & 21N 15E & 2000 June 8 & 12:00 & 44 & 2000
   June 8 & 14:00 & 6 & 120 \cr

16 & 2000 Sep 16 & 05:00 & \nodata & 2000 Sep 18 & 01:00 & 44 & 2000
   Sep 17 & 04:00 & 7 & 90 \cr

17 & 2003 Oct 28 & 11:00 & 16S 08E & 2003 Oct 29 & 08:00 & 21 & 2003
   Oct 29 & 12:00 & 21 & 150 \cr \hline
   
\end{tabular*}
%\enddata
\tablecomments{CME-ICME pairs and associated Forbush events. Under the
CME header are listed the date and time of the CME's first appearance
in the \emph{LASCO} chronograph aboard \emph{SOHO} and the
heliographic latitude and longitude of the associated source regions.  
Under the ICME header is the ICME's time of arrival at the L1
Lagrange point (Events 8--17) or when this is unknown, the time of
geomagnetic storm sudden commencement is used, which is a proxy for
the shock passage (Events 1--7).  Net transit time, $t_{transit}$, is
simply the difference of the first appearance and arrival times listed
previously. CME and ICME arrival times are taken from listings by
\emph{Gopalswamy} [2001], \emph{Cho et al.} [2003], and \emph{Cane et
al.} [1996], or in the case of Event 11 from \emph{Skoug et al.}
[2004]. Under the Forbush Event header is the onset time of the second
step Forbush decrease at Earth, a proxy for the ICME arrival, and our
measurements of the events' depths and recovery times, $t_{recov}$, at 1 AU using neutron fluxes measured by the ground-based Moscow
Neutron Monitor.}
\end{table*}
\end{article}

\end{document}